\documentclass[twocolumn]{jpsj3}
\usepackage{txfonts}

\def\Z{\mathbb{Z}}

\def\vect#1{\mbox{\boldmath $#1$}}

\title{Half-Quantum Vortices in Thin Film of Superfluid $^3$He}

\author{
Kenji Kondo$^1$, Tetsuo Ohmi$^2$, Mikio Nakahara$^{1,2}$, Takuto Kawakami$^3$, 
Yasumasa Tsutsumi$^4$ and Kazushige Machida$^3$
}
\inst{\address{
$^1$Department of Physics, Kinki University, Higashi-Osaka, 577-8502, Japan\\
$^2$Research Center for Quantum Computing,\\ 
Interdisciplinary Graduate School of Science and Engineering, 
Kinki University, 
Higashi-Osaka, 577-8502, Japan\\
$^3$Department of Physics, Okayama University, Okayama 700-8530,
Japan.\\
$^4$Condensed Matter Theory Laboratory, RIKEN, Wako, Saitama 351-0198, Japan}
}
\abst{
Stability of a half-quantum vortex (HQV) in superfluid $^3$He has been 
discussed recently by Kawakami, Tsutsumi and Machida in
Phys. Rev. B {\bf 79}, 092506 (2009). We further extend this work here and 
consider the A$_2$ phase of superfluid $^3$He confined in thin slab geometry 
and analyze the HQV realized in this setting. Solutions of HQV and singly 
quantized singular vortex are evaluated numerically by solving the 
Ginzburg-Landau (GL) equation and respective first critical angular velocities 
are obtained by employing these solutions. 
We show that the HQV in the A$_2$ phase is stable near the boundary between
the A$_2$ and A$_1$ phases. 
It is found that temperature and magnetic field must be fixed first in the 
stable region and 
subsequently the angular velocity of the system should be increased from zero 
to a sufficiently large value to create a HQV with sufficiently large 
probability. A HQV does not form if the system starts with a fixed angular 
velocity and subsequently the temperature is lowered down to the A$_2$ phase. 
It is estimated that the external magnetic field with strength on the order 
of 1 T
is required to have a sufficiently large domain in the temperature-magnetic 
field phase diagram to have a stable HQV.}

\kword{Half-quantum vortex, Singular vortex, First critical angular velocity}

\begin{document}

\maketitle

%%%%%%%%%%%%%%%%%%%%%%%%%%%%%%
\section{Introduction}
%%%%%%%%%%%%%%%%%%%%%%%%%%%%%%

A half-quantum vortex (HQV) in the A phase of superfluid $^3$He
has been proposed first by Volovik and Mineev in 1976 \cite{vm}.
They fully utilized the peculiar structure of the order parameter
manifold $M = S^2 \times SO(3)/\Z_2$ of the A phase, where $S^2$ is a manifold
where the unit magnetic vector $\hat{\vect{d}}$ resides, while  
$SO(3)$ is a group manifold representing the orbital degrees of 
freedom $\hat{\vect{m}}+i \hat{\vect{n}}$. 
They are intertwined by the $\Z_2$ factor, showing the
uniqueness of the magnetic part and orbital part are required 
only up to the sign flip; pairs $(\hat{\vect{d}}, 
\hat{\vect{m}}+i \hat{\vect{n}})$ and
$(-\hat{\vect{d}}, -(\hat{\vect{m}}+i \hat{\vect{n}}))$
define the same ordered state.

Since then, there have been several theoretical \cite{cb,sv,vl} 
as well as experimental \cite{hn,i}
works devoted to the stability of a HQV in a thin film of 
superfluid $^3$He-A. Recently, Kawakami, Tsutsumi and
Machida
published papers studying the Majorana modes trapped in a HQV
\cite{ktm,ktm2}, inspired by the NMR experiment conducted by Yamashita
{\it et al} \cite{y}. 
%Their analysis shows that a HQV is stablized when the 
%system is rotated with a high angular velocity on the order of 100~rad/s. 
They show that a HQV is stabilized when the system is rotated and
obtain the phase diagram in rotation velocity-system size space.
To date, however, the existence of a HQV has not been demonstrated 
in superfluid $^3$He yet in spite of intensive challenges. 

In the present paper, we further pursue their work and
show that a stable HQV exists in the 
A$_2$ phase near the boundary between the A$_1$ and the A$_2$ phases. 
The stability domain of a HQV has the temperature range
comparable to that of the A$_1$ phase if the exernal
magnetic field has a strength on the order of 1~T.

The rest of the paper is organized as follows.
In Sec. II, we introduce the Ginzburg-Landau free energy
of superfluid $^3$He to establish our notation and convention.
The order parameter describing a HQV is introduced in Sec.~III.
The critical angular velocity is defined in Sec.~IV for a
HQV and a singular vortex. Section~V is devoted to numerical
analysis. The order parameter profiles and comparison between
the formation energy of a HQV and a singular vortex are given.
Section~VI concludes this paper.

\section{Ginzburg-Landau Free Energy}

Let $A_{\alpha i}$ be the order parameter of a superfluid phase of $^3$He,
where $\alpha$ is the spinor index while $i$ is the orbital index.
Then the bulk free energy in the Ginzburg-Landau expansion 
takes the form
\begin{align}
F_B = & -\alpha A^* _{\alpha i} A _{\alpha i} +
\beta _1 A^* _{\alpha i} A^* _{\alpha i} A _{\beta j} A _{\beta j}  \notag\\
& +\beta _2 A^* _{\alpha i} A _{\alpha i} A^* _{\beta j} A _{\beta j} +
\beta _3 A^* _{\alpha i} A^* _{\beta i} A _{\alpha j} A _{\beta j}  \notag \\ 
& + \beta _4 A^* _{\alpha i} A _{\beta i} A^* _{\beta j} A _{\alpha j} +
\beta _5 A^* _{\alpha i} A _{\beta i} A _{\beta j} A^* _{\alpha j},
\end{align}
where the coefficient $\alpha$ of the second order term has a 
temperature dependence
$\alpha = \alpha' t$ with a constant $\alpha'$ and 
$t = 1 - T / T_c $. We take account of the effect of the 
strong coupling through the
paramagnon parameter $\delta$ in the fourth order terms $\beta_i$ as
\begin{align}
&\beta_1 = -(1 + 0.1 \delta)\beta_0, \quad \beta_2 = (2 + 0.2 \delta)\beta_0, \notag \\
&\beta_3 = (2 - 0.05 \delta)\beta_0, \quad 
\beta_4 = (2 - 0.55 \delta)\beta_0, \notag \\
&\beta_5 = -(2 + 0.7 \delta)\beta_0.
\end{align}

The gradient free energy is given by
\begin{align}
F_G =& K _1 \partial _i A _{\alpha j} \partial _i A _{\alpha j} ^* +
K _2 \partial _i A _{\alpha i} \partial _j A _{\alpha j} ^* \notag\\
&+
K _3 \partial _i A _{\alpha j} \partial _j A _{\alpha i} ^*.
\end{align}
The coefficients $K_i$ satisfy
\begin{align}\label{eq:wck}
K _1 = K _2 = K _3 = K
\end{align}
in the weak coupling limit. We emply the relation (\ref{eq:wck})
in the rest of this paper for simplicity. As a result, the coherence
length is uniquely defined as
\begin{align}
\xi(t) = \sqrt{\frac{K }{ \alpha}} = \sqrt{\frac{K}{\alpha'}} \frac{1}{
\sqrt{t}}.
\end{align}

In the following, we take superfluid $^3$He confined between two
parallel plates, the distance of which is less than the dipole coherence
length. The superfluid is rotated around the $z$-axis, which is perpendicular
to the plates, and subject to a strong magnetic field along the $z$-axis
so that the superfluid is in the A$_2$ phase. The parallel plates introduces
the boundary condition such that the $\hat{\vect{l}}$ vector is
perpendicular to the plates at the boundary. 
As a result, the orbital states of the
order parameter are restricted to $l_z = 1$ or $l_z=-1$ at the boundary.
A strong magnetic field along the $z$-axis aligns the $\hat{\vect{d}}$
vector in the $xy$-plane.

It turns out to be convenient in this setting to change the basis of 
%orbital part of 
the order parameter from $(x,y,z)$ to $(1,0,-1)$, 
characterizing the $z$-component of the angular momentum.
These two sets of basis vectors are related as
\begin{align}
\hat{\vect{e}} _\pm = \mp \frac{1}{\sqrt{2}} \left( \hat{\vect{e}}_x \pm i 
\hat{\vect{e}}_y \right),
\quad \hat{\vect{e}}_0 =\hat{\vect{e}}_z.
\end{align}
From now on, we change the notation of the order parameter from $A_{\alpha i}$
with respect to the $\hat{\vect{e}}_i=(\hat{\vect{e}}_x,\hat{\vect{e}}_y, 
\hat{\vect{e}}_z)$ basis to 
$A_{\mu \nu}$ with respect to $\hat{\vect{e}}_\nu =(\hat{\vect{e}}_{-},
\hat{\vect{e}}_0,
\hat{\vect{e}}_+)$ basis.

The magnetic field coupled to the superfluid changes the second order
term in $F_B$ to
\begin{align}
- \sum _{\nu = \pm} \left[(\alpha + \eta H) A_{+ \nu} ^* A_{+ \nu} + (\alpha - \eta H) A_{- \nu} ^* A_{- \nu}\right] \notag \\
=- \sum _{\nu = \pm} \alpha' t[(1 + \hat{h}) A_{+ \nu} ^* A_{+ \nu} + (1 - \hat{h}) A_{- \nu} ^* A_{- \nu}],
\end{align}
where $\hat{h} = h / t$. The variable $h = \eta H / \alpha'$
is a dimensionless parameter corresponding to the magnetic
field strength. The parameter $\eta$ is a constant
yielding coupling between $H$ and the condensate.

Let us analyze a uniform superfluind in the A$_2$ phase with $l_z=1$
by employing these free energies. Since $A_{+-} = A_{--} = 0$
for this state, the bulk free energy reduces to
\begin{align}
F_B = &- \alpha' t (1 + \hat{h}) |A _{++}|^2 + \beta_{24} |A _{++}|^4\notag\\
& - \alpha' t (1 - \hat{h}) |A _{-+}|^2 + \beta_{24} |A _{-+}|^4  \notag \\
&+ 2(\beta_{24} + 2\beta_{5}) |A _{++}|^2 |A _{-+}|^2,
\end{align}
where $\beta _{24} \equiv \beta_{2} + \beta_{4}=(4-0.35 \delta) \beta_0,
\ \beta_{24} + 2 \beta_{5} = -1.75 \delta \beta_0$.
In the weak coupling limit $\delta = 0$, we obtain $\beta_{24} + 2 
\beta_{5} = 0$
and as a result $A_{++}$ and $A_{-+}$ decouple. We also note that in case 
$\delta > 0$, we obtain $\beta_{24} + 2 \beta_{5} < 0$ and
hence the coupling between $A_{++}$ and $A_{-+}$ is attractive.

We introduce the following rescaling of physical
quantities to simplify the notations;
\begin{align}
& \beta _i \rightarrow \beta _0 \beta _i, \quad
|A_{++}|^2 , |A_{-+}|^2 \rightarrow \frac{\alpha ^{\prime} t}{\beta _0}|A_{++}|^2 , \frac{\alpha ^{\prime} t}{\beta _0}|A_{-+}|^2, \notag\\
& F_B \rightarrow \frac{ (\alpha ^{\prime} t)^2}{\beta _0} F_B.
\end{align}
Then the bulk free energy becomes
\begin{align}\label{eq:bfe}
F_B =& - (1 + \hat{h}) |A _{++}|^2 + \beta_{24} |A _{++}|^4\notag\\
&  - (1 - \hat{h}) |A _{-+}|^2 + \beta_{24} |A _{-+}|^4  \notag \\
&+ 2(\beta_{24} + 2\beta_{5}) |A _{++}|^2 |A _{-+}|^2.
\end{align}
The bulk order parameter is fixed by
minimizing Eq.~(\ref{eq:bfe}) with respect to $|A_{++}|^2$ and $|A_{-+}|^2$ 
as
\begin{align}
|A_{++}|^2 &= \frac{(1 + \hat{h}) \beta_{24} - (1 - \hat{h})(\beta_{24} + 2\beta_{5})}{-8\beta _5(\beta_{24} + \beta_{5})} \notag \\
|A_{-+}|^2 &= \frac{(1 - \hat{h}) \beta_{24} - (1 + \hat{h})(\beta_{24} + 2\beta_{5})}{-8\beta _5(\beta_{24} + \beta_{5})}.
\end{align}
We take, without loss of generality, the following sign convention
\begin{align}
A_{++} &= A ^{(0)} _{++} = -\sqrt{\frac{(1 + \hat{h}) \beta_{24} - (1 - \hat{h})(\beta_{24} + 2\beta_{5})}{-8\beta _5(\beta_{24} + \beta_{5})}} \notag \\
A_{-+} &= A ^{(0)} _{-+} = \sqrt{\frac{(1 - \hat{h}) \beta_{24} - (1 + \hat{h})(\beta_{24} + 2\beta_{5})}{-8\beta _5(\beta_{24} + \beta_{5})}}.
\end{align}
This choice gives the $\hat{\vect{d}}$-vector parallel to the $x$-axis
in the A phase resulting in the limit $\hat{h} \to 0$. 

The coherence lengths of $A_{++}$ and $A_{-+}$ in the presence of 
$\hat{h} \neq 0$ are 
$\xi_{+} = 1 / \sqrt{(1 + \hat{h})}$ and $\xi_{-} = 1 / \sqrt{(1 - \hat{h})}$
and they satisfy inequalities
\begin{align}
\xi _+ < \xi < \xi _-.
\end{align}

We now look at the gradient energy $F_G$. Consider a vortex along the $z$-axis
and assume the order parameter is translationally invariant along this
axis. Let us introduce the cylindrical coordinates $(r,\varphi,z)$ 
assuming $\partial/\partial z$ is a null operator. Let 
$n_{\mu \nu} \in \mathbb{Z}$ 
be the quantum number of the component $A_{\mu \nu}$
and write it as
\begin{align}
A _{\mu \nu} = C_{\mu \nu} (r) e^{i n_{\mu \nu} \varphi}.
\end{align}
The gradient term introduces the coupling between the orbital components
$+$ and $-$, namely the coupling between $A_{\mu +}$ and $A_{\mu -}$.
The quantum numbers $n_{\mu \pm}$ must satisfy the condition
\begin{align}\label{eq:2qcond}
n_{\mu -} = n_{\mu +} + 2
\end{align}
for the vortex to be cylidrically symmetric around the $z$-axis.
If this is the case, the gradient energy takes the form
\begin{align}\label{eq:2grade}
F_G = &\sum _{\mu , \nu , \nu ^\prime} \left[ \nu \frac{\partial}{\partial r} C_{\mu \nu} - \frac{n_{\mu \nu}}{r} C_{\mu \nu} \right]
\left[ \nu ^\prime \frac{\partial}{\partial r} C_{\mu \nu ^\prime} - \frac{n_{\mu \nu ^\prime}}{r} C_{\mu \nu ^\prime} \right]\notag
\\
&+ \sum _{\mu , \nu} \left[ \left( \frac{\partial}{\partial r} C_{\mu \nu} \right) ^2 + \frac{n ^2 _{\mu \nu}}{r ^2} \left( C_{\mu \nu} \right) ^2 \right],
\end{align}
where the rescalings
$r \rightarrow \xi r$ and $F_G \rightarrow \left[(\alpha' t)^2 / \beta _0 \right] F_G$
have been made as before.

\section{Half-Quantum Vortex}

The order parameter of a HQV proposed by Volovik and Mineev \cite{vm}
takes the form
\begin{align}\label{eq:3vmop}
A_{\alpha i}& = \Delta _A \hat{d} _\alpha (\hat{m} + i \hat{n})_i \notag \\
&= \Delta _A e^{i \varphi /2} \left( \cos{\frac{\varphi}{2} \hat{\vect{e}}_x} + \sin{\frac{\varphi}{2} \hat{\vect{e}}_y} \right) _\alpha (\hat{\vect{e}}_x
 + i \hat{\vect{e}}_y)_i
\end{align}
in the A-phase with vanishing magentic field $H=0$,
where it is assumed that the $\hat{\vect{l}}$-vector is directed
along the $z$-axis,
while the $\hat{\vect{d}}$-vector points in the $xy$-plane. 
Equation (\ref{eq:3vmop}) is rewritten as
\begin{align}\label{eq:3hqvop}
%e^{i \varphi /2} \left( \cos{\frac{\varphi}{2} \hat{\vect{e}}_x}
% + \sin{\frac{\varphi}{2} \hat{\vect{e}}_y} \right)
%= \frac{1}{\sqrt{2}} 
\Delta_A \left( \hat{\vect{e}}_+ 
- e^{i \varphi} \hat{\vect{e}}_- \right)_{\alpha} \hat{\vect{e}}_{+ i}
\end{align}
This shows that the order parameter of the HQV represented in the $(1,0,-1)$
basis has a non-vanishing winding number only in the component 
$\hat{\vect{e}}_-$. Similarly, there is an order parameter of a HQV, in which
only the component $\hat{\vect{e}}_+$ has a non-vanishing winding number.

By considering the condition (\ref{eq:2qcond}), the order parameter
(\ref{eq:3hqvop}) yields a vortex with quantum numbers
\begin{align}
\left( (n_{++},n_{+-}),( n_{-+},n_{--}) \right) = \left( (0,2),(1,3) \right).
\end{align}
We call this vortex as a vortex of type $(0,1)$ to distinguish it from other
types of vortices introduced in the following. When the superfluid is rotated
in the opposite sense, the resulting vortex has an order parameter
in which $e^{i \varphi}$ is replaced by $e^{-i \varphi}$ in
Eq.~(\ref{eq:3hqvop}), which will be called a vortex of type $(0,-1)$ having
quantum numbers
\begin{align}
\left( (n_{++},n_{+-}),( n_{-+},n_{--}) \right) = \left( (0,2),(-1,1) \right).
\end{align}
It is important to realize that the structure of a vortex of type $(0,1)$, 
obtained by rotating the superfluid in the positive sense with respect to the 
$\hat{\vect{l}}$-vector, is different from that of a vortex of type $(0,-1)$
obtained by rotating the superfluid in the oppsite direction. The condensate
with orbital angular momentum spontaneously breaks the rotational invariance 
and hence the clockwise rotation and anticlockwise rotation are not
mirror reflections of each other.

\section{First Critical Angular Velocity}

Let $R$ be the radius of a cylindrical container and $\Omega$ be
the angular velocity with which the cylinder rotates. Now we obtain
the condition under which a vortex stably exists at the center of the 
container. The gradient free energy in the rotating system 
is obtained by replacing the $\varphi$-derivative as
\begin{align}
\frac{\partial_\varphi}{ir} \rightarrow \frac{\partial_\varphi}{ir} - \frac{2 m}{\hbar} (\vect{\Omega} \times \vect{r})_\varphi,
\end{align}
where $m$ is the mass of a $^3$He atom. Let us first consider a HQV, 
%in which the lower creation energy component $\hat{\vect{e}}_-$ has 
in which the component $\hat{\vect{e}}_-$ with lower creation energy
has a non-vanishing quantum number $n_{-+}$. 
There are two terms of the form
%It follows from the gradient
%energy (\ref{eq:2grade}) that the derivative term reduces to
\begin{align}
%2 \left|\frac{ \partial_\varphi A_{-+}}{r} \right| ^2 \rightarrow 
\left( \frac{1}{r} - \frac{2 m}{\hbar} \Omega r \right) ^2 C_{-+} (r)^2
\end{align}
in the gradient energy (\ref{eq:2grade}).
The coefficient of a term linear in $\Omega$ is nothing but
the angular momentum and the total angular momentum of the system 
is found to be
\begin{align}
L^{(-)} = 2 \times 4 \pi \frac{2 m}{\hbar} \int_0^R r {\rm d} r C_{-+} (r)^2
= 4 \pi \frac{2 m}{\hbar} \left( A_{-+} ^{(0)} \right) ^2 R ^2,
\end{align}
where we noted that the contribution of the vortex core
to the total angular momentum is negligible. 

The vortex formation energy measured with respect to the uniform bulk energy
$F_0$ is evaluated as
\begin{align}
F_{{\rm vor}} ^{(-)} =   2 \pi \int r {\rm d} r (F - F_0)
= 4 \pi \left( A_{-+} ^{(0)} \right) ^2 (\ln R + C_-).
\end{align}
The parameter $C_-$ will be evaluated numerically later.
The first critical angular velocity for a formation of a vortex
with $\hat{\vect{e}}_-$ spin component, namely a vortex in the spin component
$\downarrow \downarrow$ is obtained by solving
\begin{align}
F_{{\rm vor}} - \Omega L = 0
\end{align}
as
\begin{align}
\Omega_{c} ^{(-)} = \frac{\hbar}{2 m} \frac{F_{{\rm vor}} ^{(-)}}{\left( A_{-+} ^{(0)} \right) ^2 R ^2}
= \frac{\hbar}{2 m R ^2} (\ln R + C _-)
\end{align}
An angular velocity will be scaled by $\hbar / 2 m R ^2$ from now on.
As a result, the crtical angular velocity is written as
\begin{align}\label{eq:1cavhqv}
\Omega_{c} ^{(-)} = \ln R + C _-.
\end{align}

A singular votex (SV) with a winding number 1 is obtained by setting
the quantum numbers of $A_{\mu \nu}$ to
\begin{align}
\left( (n_{++},n_{+-}),( n_{-+},n_{--}) \right) = \left( (1,3),(1,3) \right).
\end{align}
The total angular momentum of a SV is
\begin{align}
L^{(s)} = &2 \times 4 \pi \frac{2 m}{\hbar} \int_0^R r {\rm d} r \left[ C_{++} (r)^2 + C_{-+} (r)^2 \right]\notag \\
=& 4 \pi \frac{2 m}{\hbar} \left[ \left( A_{++} ^{(0)} \right) ^2 + \left( A_{-+} ^{(0)} \right) ^2 \right] R ^2.
\end{align}
The formation energy of a SV is 
\begin{align}
F_{{\rm vor}} ^{(s)} = 4 \pi \left[ \left( A_{++} ^{(0)} \right) ^2 + \left( A_{-+} ^{(0)} \right) ^2 \right] (\ln R + C_s)
\end{align}
and the first critical angular velocity is
\begin{align}\label{eq:1cavsv}
\Omega_{c} ^{(s)} = \ln R + C _s
\end{align}
in the dimensionless form.

Whether a HQV forms or a SV forms as the angular velocity increases
depends on the magnitude relation between $\Omega _c ^{(-)}$ and
$\Omega _c ^{(s)}$. Let us consider the case $\delta = 0$ and $\hat{h} > 0$
to begin with. It follows from the inequality $\xi _+ < \xi _-$ that
a vortex in $A_{-+}$ is energetically favorable than that in $A_{++}$
and it is expected that $\Omega _c ^{(-)} < \Omega _c ^{(s)}$ is satisfied.
In case $\delta > 0$ and $\hat{h} = 0$, the coupling between  
$A_{++}$ and $A_{-+}$ is attractive and a SV is expected to be
favorable compared to a HQV. This is because low magnitude $A_{++}$ and 
$A_{-+}$ overlap at the common vortex core in a SV while they do not
in a HQV, thus gaining more negative energy for the former.
Then an inequality $\Omega _c ^{(-)} > \Omega _c ^{(s)}$ is 
expected to be satisfied. 

It is expected from the above arguments that
a SV is formed first as the angular velocity is raised from zero
when $\delta > 0$ and $\hat{h}$ is small. 
When the external magnetic field is strong enough, in
contrast, there is a region in the temperature-angular velocity domain
in which a HQV is formed first. These statements will be verified
numerically in the next section.

\section{Numerical Analysis}

We have solved the Ginzburg-Landau equation with respect to
$(C_{++} (r) , C_{+-} (r) , C_{-+} (r) , C_{--} (r))$ numerically. 
Four choices of the quantum numbers are considered;
\begin{enumerate}
\item[(a)] A HQV $(0,1)$ with 
$\left( (n_{++},n_{+-}),( n_{-+},n_{--}) \right) = \left( (0,2),(1,3) \right)$,
\item[(b)] A SV $(1)$ with 
$\left( (n_{++},n_{+-}),( n_{-+},n_{--}) \right) = \left( (1,3),(1,3) \right)$,
\end{enumerate}
both with $\Omega >0$ and
\begin{enumerate}
\item[(c)] A HQV $(0,-1)$ with 
$\left( (n_{++},n_{+-}),( n_{-+},n_{--}) \right) = \left( (0,2),(-1,1) \right)$,
\item[(d)] A SV $(-1)$ with 
$\left( (n_{++},n_{+-}),( n_{-+},n_{--}) \right) =
 \left( (-1,1),(-1,1) \right)$,
\end{enumerate}
both with $\Omega < 0$. The parameters $\hat{h}$ and $\delta$
are changed from 0.0 to 0.5 with a step 0.1.
The boundary condition at $r=R$ does not affect the formation energy since 
$R \gg 1$ is assumed. We take the boundary condition
\begin{align}
&\left( A_{++} (R) , A_{+-} (R) ,  A_{-+} (R) , A_{--} (R) \right)\notag \\
& = \left( A_{++} ^{(0)} , 0 , A_{-+} ^{(0)} , 0 \right),
\end{align}
which corresponds to a vortex embedded in a uniform $\hat{\vect{l}}$
texture with $l_z=+1$.
\begin{figure}
\begin{center}
\includegraphics[width=7cm]{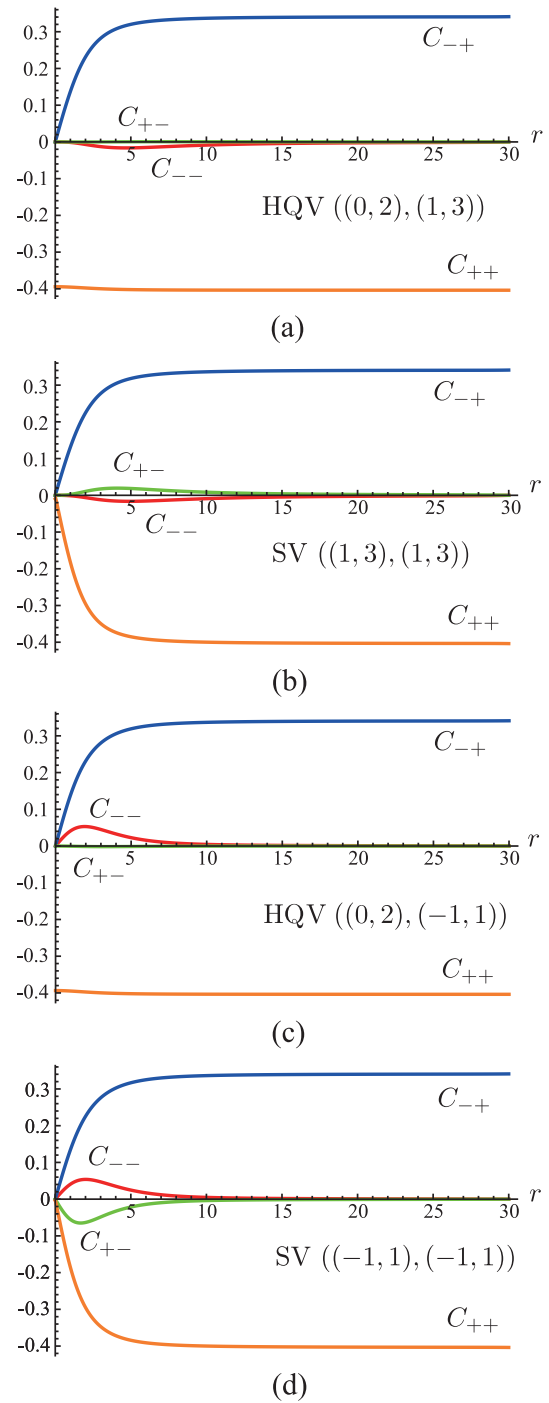}
\end{center}
\caption{(Color online) 
Vortex solutions of the Ginzburg-Landau equation. 
Order parameters $C_{\pm\pm}(r)$
are shown for $\delta=0.2, \hat{h} = 0.2$ and $R=30$. Quantum numbers
$((n_{++},n_{+-}),(n_{-+},n_{--}))$
are (a) $((0,2),(1,3))$, corresponding to a HQV with $\Omega>0$ (b) 
$((1,3),(1,3))$, corresponding to a SV with $\Omega>0$ (c) $(0,2),(-1,1))$,
corresponding to a HQV with $\Omega<0$ and
(d) $((-1,1),(-1,1))$ corresponding to a SV with $\Omega <0$.}
\label{fig:1}
\end{figure}

The order parameter profiles for HQV~$(0,1)$, SV~$(1)$,
HQV~$(0,-1)$ and SV~$(-1)$ with 
$\delta = 0.2, \hat{h} = 0.2$, and $R = 30$
are shown in Fig.~\ref{fig:1}.
%Those for  with the same parameters
%are also shown in Fig.~\ref{fig:2}.

Next the first critical angular velocities are obtained by
evaluating the free energies of HQV's and SV's with our numerical 
solutions and then employing Eqs.~(\ref{eq:1cavhqv}) and (\ref{eq:1cavsv}). 
Furthermore, we repeat the same calculation with $R=40$ and $50$ and
fit the first critical angular velocities thus obtained
with a function $\Omega_c=A \ln R +C$, $A$ and $C$ being constants.
The result shows that $A$ is in fact 1 with a good precision, 
% with a good precision
as expected,
and we have determined the $\hat{h}$- and $\delta$-dependences of 
$C _-$ and $C _s$, the $C$ value of a HQV $(0,1)$ and a SV $(1)$,
respectively.
\begin{figure}
\begin{center}
\includegraphics[width=7cm]{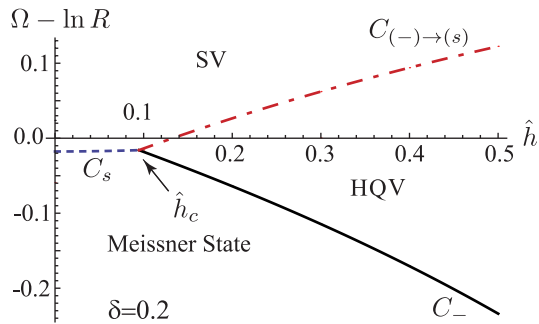}
\end{center}
\caption{(Color online) 
HQV-stable region for $\Omega>0$. The solid line is the boundary
between the Meissner state without a vortex and a HQV while the dotted
line is the boundary between the Meissner state and a SV. For sufficiently
large $\Omega$, transition from a SV to a HQV takes place as 
$\hat{h}$ is increased and the dashed line is crossed. The area
bounded by the solid line and the dashed line is the region where
a HQV has the least energy. 
%It shows that there exists a frequency interval in which a HQV is stable 
%for any $\hat{h}$ when $\delta =0$, 
A HQV-stable angular velocity region exists only for 
$\hat{h}> \hat{h}_c(\delta)$ when $\delta >0$, while it exists
for any $\hat{h}$ when $\delta =0$. The
intersecting point of the solid, the dashed and the dotted
lines gives $\hat{h}_c(\delta)$.}
\end{figure}
Figure~2 shows $C_- = \Omega _c ^{(s)} - \ln R$ and 
$C_s = \Omega _c ^{(s)} - \ln R$ for cases (a) and (b) 
as functions of $\hat{h}$ for $\delta = 0.2$.
The dimensionless magnetic field $\hat{h}$ also takes values
$\hat{h}= 0.0, 0.1, \ldots, 0.5$. The result shows that,
for $\delta \neq 0$, there
exists $\hat{h} _c$ at which the inequality
$\Omega _c ^{(-)} > \Omega _c ^{(s)}$ 
flips to $\Omega _c ^{(-)} < \Omega _c ^{(s)}$ as $\hat{h}$ is increased.
The critical magnetic field $\hat{h}_c$ vanishes for $\delta =0$,
showing there is a range of $\Omega$ in which a HQV is
stable for any $\hat{h} >0$.
The first critical angular velocity $\hat{h} _c$ has been estimated
by finding the intersection of numerically interpolated curves 
$\Omega _c ^{(-)}$ and $\Omega _c ^{(s)}$ as functions of $\hat{h}$.
\begin{figure}
\begin{center}
\includegraphics[width=7cm]{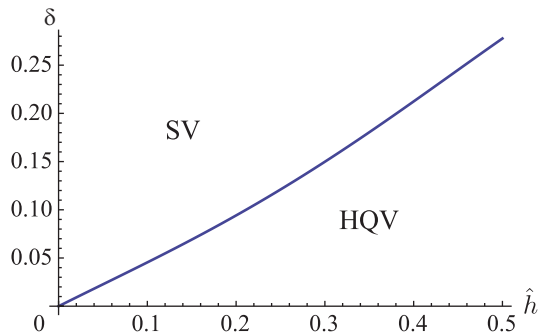}
\end{center}
\caption{(Color online) 
HQV-stable region in the $\hat{h}$-$\delta$ domain for $\Omega>0$.
The boundary corresponds to $\hat{h}_c(\delta)$.
Here HQV denotes the parameter region in the $\hat{h}$-$\delta$
plane where there exist $\Omega>0$
for which a HQV has lower energy than a SV.
%means those area in the $\hat{h}$-$\delta$,
%where a HQV-stable angluar velocity region exists. 
There is
essentially no difference in the graph for $\Omega<0$
in the $\hat{h}$-$\delta$ plane.}
\end{figure}
Figure~3 depicts the $\delta$-dependence of $\hat{h} _c$ thus obtained.
In case $\Omega _c ^{(-)} < \Omega _c ^{(s)}$, we reach
a region in the $\Omega$ axis in which a SV is stablized
if the angular velocity is further increased beyond $\Omega _c ^{(-)}$.
The boundary between two stability regions along the $\Omega$-axis is
found from
\begin{align}
F_{{\rm vor}} ^{(-)} - \Omega L^{(-)} = F_{{\rm vor}} ^{(s)} - \Omega L^{(s)}.
\end{align}
Figure~2 also shows $C_ {(-) \rightarrow (s)} = \Omega_c^{(-) \rightarrow (s)} - \ln R$. The region in which a HQV is stable is bounded by
two curves $\Omega _c ^{(-) \rightarrow (s)}$ and $\Omega _c ^{(-)}$ in
Fig.~2.

\begin{figure}
\begin{center}
\includegraphics[width=7cm]{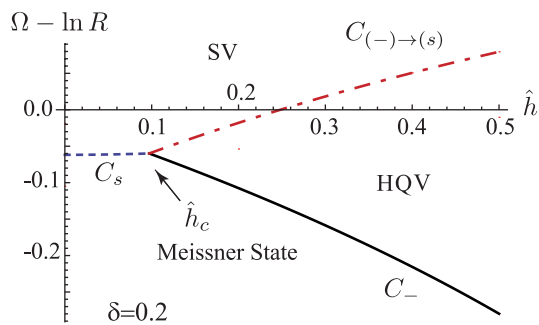}
\end{center}
\caption{(Color online) HQV-stable region for $\Omega < 0$. The solid line,
the dashed line and the dotted line denote the same
boundaries as in Fig.~2.}
\end{figure}
Next, the $\hat{h}$-dependences of $C _-$ and $C _s$ for cases (c) and
(d), respectively, are depicted in Fig.~4. They are different from
those of cases (a) and (c), reflecting upon the difference in
the vortex structures for $\Omega >0$ and $\Omega<0$. 
Figure~4 also shows $\Omega _c ^{(-) \rightarrow (s)}$,
similarly to Fig.~2. A HQV is stable in the region bounded by two curves
$\Omega _c ^{(-) \rightarrow (s)}$ and $\Omega _c ^{(-)}$. 
Let us evaluate the critical value $\hat{h} _c$,
at which the HQV-stable region appears. The critical value will turn out 
to be almost
the same as that with $\Omega > 0$, in contrast with $C _-$ and $C _s$.
The region in the $t \delta$-plane (the 
temperature-pressure plane) where a HQV is stable
is obtained from the $\delta$-dependence of $\hat{h} _c$.
The phase diagram for a given $h = \eta H / \alpha'$ is shown in
Fig.~5, where $t = h / \hat{h}$ has been used. 
\begin{figure}
\begin{center}
\includegraphics[width=7cm]{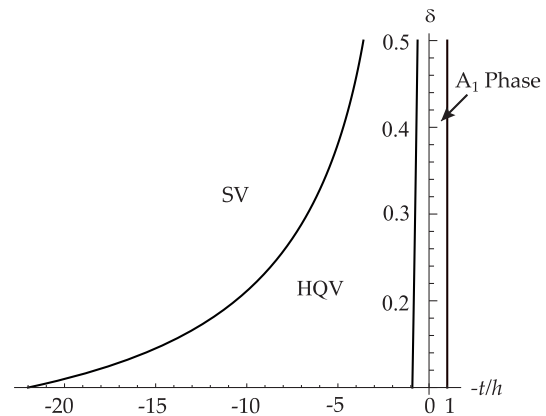}
\end{center}
\caption{(Color online) 
HQV-stable region in the $t$-$\delta$ plane. The HQV-stable region
is determined by the $\delta$-dependence of $\hat{h}_c$.
For comparison, the A$_1$ phase, given $h=\eta H/\alpha'$,
is also shown here. The origin of the horizontal axis
corresponds to $T=T_c$ at $H=0$. The critical temperature $T_c$ 
of the A$_1$-phase is 1 in the present scaling.
The $\delta$-dependence of $\eta$ and
$T_c$ at $H=0$ is ignored and, hence, comparison of the 
phases with different $\delta$ should not be taken seriously.
In spite of this, comparison of the widths of the A$_1$ phase
and the HQV-stable region is meaningful. It shows that the
width of the HQV-stable region has the width of the same
order of that of the A$_1$ phase. The
width of the former increases as $\delta$ is lowered 
(low pressure region).}
\end{figure}
The A$_1$ phase is also shown in Fig.~5 for comparison.
Although we have ignored the $\delta$-dependences of $\eta$
and the superfluid transition temperature $T_c$ at $H=0$,
the comparison between the width of the A$_1$ phase and that
of the HQV-stable region is meaningful for a fixed $\delta$.
Figure~5 shows that the width of the HQV-stable region along the
$t/h$-axis is comparable to
that of the A$_1$ phase. Morevoer the former 
increases compared with the latter for small $\delta$ (low pressure)
region. 

\section{Conclusion and Discussion}

We have obtained conditions with which a half-quantum vortex stably exists
and have shown that the stability region of a HQV has a comparable range
to that of the A$_1$ phase along the $t/h$-axis. 

To obtain the HQV, the temperature
and the magnetic field must be fixed in the HQV-stable region with no
rotation first and 
subsequently the angular velocity must be increased beyond the critical
angular velocity. 
The opposite scenario, in which the system is rotated beyond 
the critical angular velocity first and then the temperature is lowered to
form the A$_2$ phase through the A$_1$ phase, does not lead to a HQV formation.
This is because a sigular vortex forms in the $\uparrow \uparrow$-component
while the system is in the A$_1$ phase and it is impossible to eliminate this
singular vortex after the system reaches the A$_2$ phase.
We believe that a magnetic field on the order of 1~T is required
to have a large enough HQV-stable region comparable to that of the A$_1$
phase. 

It is desirable to observe the direct NMR signal from a HQV for
its detection. Nonetheless, direct observation can be rather 
challenging in the presence of a strong magnetic field.  
Note, however, that, when HQV formation takes place,
there are two transitions associated with vortex
formation in the vicinities of 
$\Omega _c ^{(-)}$ and $\Omega _c ^{(-) \rightarrow (s)}$ as the angular
velocity is increased. We expect these two transtions can be experimentally
observable by one way or another.

\section*{Acknowledgements}

TO and MN are
supported by ``Open Research Center'' Project for Private Universities;
matching fund subsidy from 
the Ministry of Education, Culture, Sports, Science and Technology (MEXT), 
Japan.
MN is also supported by the ``Topological Quantum Phenomena'' 
(No. 22103003) Grant-in Aid for Scientific Research on Innovative Areas from 
MEXT, Japan.
KM is supported by Grant-in-Aid for Scientific Research (B) (No. 21340103).
TK is supported by 
Grant-in-Aid for JSPS Fellows (No. 2200247703).

%--------------------------%

\end{document}